\documentstyle[12pt]{article}
\newcommand{\beq}{\begin{equation}}
\newcommand{\eeq}{\end{equation}}
\newcommand{\be}{\begin{equation}}
\newcommand{\ee}{\end{equation}}
 \newcommand{\bi}{\bibitem}

\begin{document}
\begin{center}
\vglue .06in
\begin{center}
{\Large \bf {THE PROBLEM OF VACUUM ENERGY AND COSMOLOGY\\
({\it A lecture presented at the 4th Colloque Cosmologie, Paris, June, 1997)}
  }}
\end{center}
\bigskip
{\bf A.D. Dolgov}
 \\[.05in]
{\it{Teoretisk Astrofysik Center\\
 Juliane Maries Vej 30, DK-2100, Copenhagen, Denmark
\footnote{Also: ITEP, Bol. Cheremushkinskaya 25, Moscow 113259, Russia.}
}}
\end{center}
\begin{abstract}
 Quantum field theory predicts that vacuum energy (or what is the same,
cosmological constant) should be 50-100 orders of magnitude larger than
the existing astronomical limit. A very brief review of possible solutions
of this problem is presented. A mechanism of adjustment of vacuum energy
down to (almost) zero by the back-reaction of massless vector or second
rank tensor fields is discussed.
\end{abstract}
\bigskip

\section{Introduction }

The problem of cosmological constant, $\Lambda$, or, what is the same, of
vacuum energy is one of the most or just the most striking problem in the
contemporary fundamental physics. This is a unique case when theoretical
expectations differ from observations at least by $10^{45}$ or maybe even by
120 (!) orders of magnitude. It is known from cosmology that vacuum energy
would influence the evolution of the universe and from the absence of the
noticeable changes with respect to the usual Friedman model one may deduce
that $|\Omega _{vac}| = |\rho_{vac} /\rho_c | \leq 1$.
Here $\rho_c =3H^2 m_{Pl}^2 /8\pi \approx 10^{-29} {\rm g/cm}^3
\approx 10^{-47} {\rm GeV}^4$ is the critical energy density.
So we can conclude that
\be
|\rho_{vac}| < 10^{-47} {\rm GeV}^4
\label{upvac}
\ee
On the other hand, quantum field theory predicts
that there are plenty of contributions into vacuum energy which are larger
than this bound, roughly speaking, by 50-100 orders of magnitude.

Sometimes astronomers put a different meaning into the words
"the problem of cosmological constant". Namely, there is a continuous
discussion, if $\Omega_{vac}$ is exactly zero (or unnoticeably small)
or it may be close to unity so that its effects
on universe evolution are essential. In particular, the relation between the
universe age, $t_U$, and the present-day value of the Hubble constant, $H_0$,
depends upon the magnitude of $\Lambda $ and the discrepancy
between the large values of $H_0$ and $t_U$ would disappear if
$\Omega _{vac} = 0.7-0.8$. With the same value of $\Omega _{vac}$ the theory
of large scale structure formation gives a better description of the data than
just with $\Lambda = 0$. (Let us note that the relation between $\rho_{vac}$
and $\Lambda$ is given by the expression (\ref{lamrho}) below.)
It is quite mysterious why the
value of $\rho_{vac}$, which remains constant in the course of the universe
expansion, is so close {\it today} to the value of the critical energy density
which falls down as $1/t^2$ (in cosmologies with $\Omega_{tot} =1$). It
adds up to the other two mysteries: why the energy density of baryons, which
contributes into $\Omega$ at a per cent level, and energy densities of hot and
cold dark matter (if both or any one of the latter exist) all are close to each
other within the order of magnitude, though they seem to be unrelated and could
easily differ by several (many?) orders of magnitude.

We will concentrate on the first problem, namely why vacuum energy is tiny
on the scale of elementary particle physics despite all huge contributions
(see below). The other problem, if vacuum energy is cosmologically noticeable,
is not directly addressed here but it is quite possible that the solution to
the first problem (which is unknown at the present day) could help to solve
the second one too.

\section{History of the Problem}

Cosmological constant was introduced in 1918 by Einstein \cite{ae} when he
unsuccessfully tried to apply General Relativity equations to cosmology and
was disappointed to find that there were no stationary solutions. The usual
Einstein equations have the form:
\be
R_{\mu\nu} -  {1\over 2} g_{\mu\nu} R =
{8\pi T^{(mat)}_{\mu\nu}\over m_{Pl}^2 },
\label{eineq}
\ee
where the source of gravity is the energy-momentum tensor of matter,
$T^{(mat)}_{\mu\nu}$. These equations do not have static solutions for
homogeneous and isotropic distribution of matter. To overcome this, what seemed
to be a shortcoming, Einstein proposed to add an extra term into this
equations, $\Lambda g_{\mu\nu}$:
\be
R_{\mu\nu} -  {1\over 2} g_{\mu\nu} R =
{8\pi T^{(mat)}_{\mu\nu}\over m_{Pl}^2 } + \Lambda g_{\mu\nu}.
\label{einlam}
\ee
where $g_{\mu\nu}$ is the metric tensor and
$\Lambda$ is a constant which got the name cosmological constant. As it was
understood later this new term corresponds to a nonzero energy-momentum tensor
of vacuum and $\Lambda$ is related to the vacuum energy density as:
\be
\Lambda = 8\pi \rho_{vac} /m_{Pl}^2
\label{lamrho}
\ee
It can be shown that a positive $\rho_{vac}$ induces gravitational repulsion,
so that the introduction of the cosmological term may balance the gravitational
attraction of the usual matter and correspondingly may lead to
stationary solutions of eq.~(\ref{einlam}). To this end a careful tuning
of $\rho^{(mat)}$ and $\rho_{vac}$ should be arranged. What is worse,
this static solution is evidently unstable with respect to density
perturbations. After Friedman \cite{af} had found the non-stationary
cosmological solution and especially after Hubble \cite{hub}
discovered that the universe indeed expands in accordance with this solution,
Einstein very strongly objected the idea of nonzero lambda-term and
considered it as the biggest blunder of his life. In drastic contrast to that
opinion Lemaitre \cite{lem} advocated the introduction of cosmological constant
as a great discovery.

It was much later stressed by Zeldovich \cite{ybz} that quantum field theory
generically demands that cosmological constant or, let us repeat, what is the
same, vacuum energy is non-vanishing. It is very well known from quantum
mechanics that the ground state energy of harmonic oscillator is not zero
but equal to $\omega/2$. It can be understood in the following way: a
particle "siting" on the bottom of harmonic oscillator potential must have a
nonzero momentum due to uncertainty principle. Correspondingly its energy
should be nonzero. Similar phenomenon takes place in quantum field theory
because any quantized field can be represented as a collection of oscillators
with all possible frequencies. Correspondingly the ground state energy of
this system is given by the expression:
\be
\rho_{vac} = g_s\int {d^3 p  \over (2\pi)^3 }\sqrt{ p^2 + m^2} = \infty ^4
\label{rhob}
\ee
Here $m$ is the mass of the field, $g_s$ is the number of spin states of
the field and it is assumed that the field in question is a bosonic one.

One cannot live in the world with infinitely big vacuum energy, so Zeldovich
assumed that bosonic vacuum energy should be compensated by vacuum
energy of fermionic fields. Indeed, vacuum energy of fermions
is shifted down below zero and is given by exactly the same integral as
(\ref{rhob}) but with the opposite sign. (This is related to the condition that
bosons are quantized with commutators while fermions are quantized with
anti-commutators.) So, if there is a symmetry between bosons and fermions
such that for each bosonic state there exists a fermionic state with the
same mass and vice versa, then the energy of vacuum fluctuations of bosons and
fermions would be exactly compensated, giving zero net result. This
assertion \cite{ybz} was made before the the pioneering works on
supersymmetry \cite{gl,va,wz} were published. However, since supersymmetry is
not exact this compensation is not complete (see the next Section).

\section{Sources of Vacuum Energy}

It is not excluded experimentally that the number of fermionic and
bosonic species in Nature are the same. Moreover it is practically a
necessity, because otherwise vacuum energy density would be infinite. Still the
masses of bosons and corresponding fermions are different and, with
arbitrary relations between their masses, only the leading term, which diverges
as the fourth power of the integration limit, would be canceled out.
However in some supersymmetric theories with spontaneous symmetry breaking
there may be specific relations between masses of different fields which
ensure the compensation not only of the leading term but also quadratically
and logarithmically divergent terms.
This looks as a very strong argument in favor of such models. However the
finite terms are not compensated. Moreover in global supersymmetric theories
finite contributions into $\rho_{vac}$ {\it must} be nonzero and by the order
of magnitude they are equal to
\be
\rho_{vac}^{(susy)} \sim m^4 _{susy}
\label{rhosusy}
\ee
where $m_{susy}$ is the scale of supersymmetry breaking. It is known from
experiment that $m_{susy} \geq 100 $ GeV. Correspondingly
$\rho_{vac}^{(susy)} \geq 10^8 $ GeV, i.e. 55 orders of magnitude larger
than the permitted upper bound. In more advanced supersymmetric theories
which include gravity (the so called supergravity or local supersymmetry) the
condition of non-vanishing vacuum energy in the broken symmetry phase is not
obligatory. However, if one does not take a special care, the value of
vacuum energy in unbroken supergravity models is typically about
$m^4_{Pl} \approx 10^{76}$ GeV.
One can choose in principle the parameters in such a way that this
contribution into $\rho_{vac}$ is compensated down to zero with the
accuracy $10^{-123}$ but this demands a fantastic fine-tuning.

One more source of vacuum energy is the energy of the scalar (Higgs) field
in the theories with spontaneous symmetry breaking. The potential of such
field  is typically of the form:
\be
U(\phi ) = -m^2 \phi^2 + \lambda \phi^4
\label{uphi}
\ee
This potential has minimum at $\phi^2 = m^2/2\lambda \neq 0$. At nonzero
temperatures it acquires corrections of the form $\sim T^2 \phi^2$ so at
sufficiently high $T$ the minimum of the potential shifts to $\phi = 0$. Thus,
in the early universe the ground state in such a theory was at $\phi =0$ and
in the course of the universe cooling down a phase transition \cite{kir,kl}
took place to the state with nonzero $\phi$. The change of vacuum energy in
the course of such phase transition is $\delta \rho_{vac} = m^4/4\lambda$.
Accordingly the electroweak phase transition contributes about
$10^{10}\,{\rm GeV}^4$ into $\rho_{vac}$ and the grand unification one
gives more than  $10^{60}\, {\rm GeV}^4$.

One could argue that these phase transitions are manifestations of high energy
physics and who knows, if they existed or not. The vacuum  energy might be
adjusted in such a way that it is always zero in the broken symmetry phase
when $\phi \neq 0$. This corresponds to the choice of the potential
$U(\phi)$~(\ref{uphi}) in the form $U(\phi) = \lambda (\phi^2 - \eta^2)^2$.
So maybe these huge contributions into $\rho_{vac}$ are just products of
our imagination. However there exist some other contributions which, though
smaller than the grand unification and even the electroweak ones, are still
huge in comparison with $10^{-47} \,{\rm GeV}^4$. It is well known that
vacuum state in quantum chromodynamics (QCD) is not empty. It is filled by
non-perturbative quark (or chiral) condensate \cite{gm} and gluon
condensate \cite{svz}. The existence of these condensates is practically an
experimental fact. Successful QCD description of hadron properties is
impossible without these condensates. The vacuum energy density of the quark
condensate $\langle \bar q q \rangle$ is about $10^{-4}\,{\rm GeV}^4$ and
that of gluon condensate $\langle G_{\mu\nu}^2 \rangle$ is approximately an
order of magnitude bigger. Comparing these numbers with the upper
bound~(\ref{upvac}) we see that there {\it must} exist something which does
not know anything about quarks and gluons (this "something" is not related
to quarks and gluons by  the usual QCD interactions, otherwise it will be
observed at experiment) but still this mysterious agent is able to
compensate their vacuum energy with the fantastic accuracy of $10^{-44}$.

The problem seems to be very serious and most probably demands new physics
beyond the known standard model. An important feature that makes the solution
of the problem especially difficult, is that the looked for modification
should be done in low energy physics, corresponding to the energy scale
about $10^{-3}\, {\rm eV}$.

\section{Possible Models of "Nullification" of Vacuum Energy}

Several possible approaches to the problem of vacuum energy have been discussed
in the literature, for the review see refs. \cite{sw1,ad1}. They can be
roughly put into four different groups:
\begin{enumerate}
\item{} Modification of gravity on large scales.
\item{} Anthropic principle.
\item{} Symmetry leading to $\rho_{vac} = 0$.
\item{} Adjustment mechanism.
\end{enumerate}

A modification of gravity at large scales should be done in such a way that the
general covariance, which ensures vanishing of the graviton mass, is preserved,
energy momentum tensor is covariantly conserved, and simultaneously the vacuum
part of this tensor, which is proportional to $g_{\mu\nu}$, does not gravitate.
This is definitely not an easy thing to do. Possibly due to this reasons
there is no satisfactory model of this kind at the present time.

Anthropic principle states that the conditions in the universe must be
suitable for life, otherwise there would be no observer that could
put a question why the universe is such and not another. With cosmological
constant which is as large as predicted by natural estimates in quantum theory,
life of our type is definitely impossible. Still this point of view does
not look very appealing. The situation is similar to the one that existed
in the Friedmann cosmology before inflationary resolution of the fundamental
cosmological problems has been proposed \cite{guth}.

There is one more difficulty  in the implimenttion of the anthropic principle.
Even if we assume that it is
effective, there are no visible building blocks to achieve the necessary
compensation of vacuum energy. One can say of course that this compensation
is not achieved by a physical field but just by a subtraction constant or in
other words by a choice of the position of zero on the energy axis. In other
words it is assumed that there is some energy coming from nowhere, which
exactly cancels out all the contributions of different physical fields.
Though formally this is not excluded, it definitely does not look
beautiful.

Probably the most appealing would be a model based on a symmetry principle
which forbids a nonzero vacuum energy. Such a symmetry should connect known
fields with new unknown ones. Some of those fields should be very light
to achieve the cancellation on the scale $10^{-3}$ eV. Neither such fields
are observed, nor such a symmetry is known.

An adjustment mechanism seems to me the most promising one at the present time.
The idea is similar to the mechanism of solving the problem of natural
CP-conservation in quantum chromodynamics by the axion field~\cite{pq,sw2,wil}.
The axion potential automatically acquires a minimum at the value of the
field amplitude that cancels out the CP-odd contribution from the so called
theta-term, $\theta G \tilde G $. Similar mechanism can hopefully kill vacuum
energy. Let us assume that there is a very light or massless field coupled to
gravity in such a way that it is unstable in De Sitter background and develops
the condensate whose energy-momentum tensor is equal by magnitude and opposite
by sign to the original vacuum energy-momentum tensor. Though it looks rather
promising, it is very difficult, if possible at all, to construct a realistic
model based on this idea.  Some of the existing attempts to do that, are
discussed in the following three sections.

\section{Adjustment by a Scalar Field}

A scalar field looks the most natural for the role of the adjustment
agent. This is why the first attempts to realize the adjustment was based on
the hypothesis on a massless scalar field non-minimally coupled to
gravity \cite{ad2}:
\be
{\cal L } (\phi) = \sqrt {-g} \left[ g^{\mu\nu} \partial_\mu\phi
\partial_\nu \phi + \xi R \phi^2 \right]
\label{lphi}
\ee
Here $\xi$ is a constant and $R$ is the curvature scalar. Such non-minimal
coupling is well known in the literature. In particular the condition of
conformal invariance of a massless scalar field  demands  $\xi=1/6$.
We will consider the evolution of homogeneous (space-point independent)
field $\phi$ in the Friedmann-Robertson-Walker background with the metric:
\be
 ds^2 = dt^2 -a^2(t) d\vec r\>^2
\label{ds2}
\ee
For simplicity we assumed that it is spatially flat.

The equation of motion for massless minimally coupled ($\xi =0$) scalar
field in this metric has the form:
\be
D^2 \phi \equiv \left( \partial_t^2 - {1\over a^2(t)}\partial_j^2
+3H \partial_t \right) \phi = 0,
\label{d2phi}
\ee
where $a(t)$ is the scale factor and $H = \dot a /a$ is the Hubble parameter.
Evidently this equation has only constant solutions or solutions decreasing
in the course of the expansion.

If $\xi \neq 0$ and the product $\xi R$ is negative, then it effectively
behaves as negative mass squared and correspondingly
the state with $\phi =0$ is unstable. Now the equation takes the the form:
\be
 \left( \partial_t^2 - {1\over a^2(t)}\partial_j^2
+3H \partial_t  +\xi R \right) \phi = 0,
\label{d2phir}
\ee
In the De Sitter space-time both the curvature scalar $R=12H^2$ and the Hubble
constant $H$ (which now is a constant indeed) are time independent and it is
easy to see that for $\xi R <0$ this equation has an exponentially rising
solution, $\phi \sim \exp (c H t)$, where $c$ is a numerical coefficient.
Thus if one starts from the state dominated by vacuum energy,
$T_{\mu\nu} \approx \rho_{vac} g_{\mu\nu}$, the
universe would initially expand exponentially, $a(t) \sim \exp (Ht)$. However
fluctuations of the field $\phi$ would be unstable in this background
and very soon the amplitude of this field would become large, so its influence
on the expansion should be taken into account. One can check that
asymptotically
$\phi \sim t$ and the exponential expansion of the universe turns into the
power law one \cite{ad2}, $a(t) \sim t^\kappa$. Thus it seems that our goal
is reached. We started from the De Sitter universe and ultimately came to
the Friedmann one. However one can check that the energy-momentum
tensor of the field $\phi$ is by no means proportional to the vacuum one,
so there is no cancellation between them. The slowing down of expansion is
achieved not by killing the anti-gravitating vacuum energy but by asymptotic
cancellation of the gravitational coupling constant. Indeed, the curvature
scalar enters the Lagrangian as $(8\pi m_{Pl}^2 - \xi \phi^2)R$. It means that
the effective gravitational coupling behaves as
\be
G^{(eff)} = {1 \over  m_{Pl}^2 + |\xi| \phi^2(t) /8\pi }
\label{geff}
\ee
So with $\phi \sim t$ the gravitational coupling dies down with time as
$1/t^2$.
This is not the solution that we looked for. Though the example itself is
rather
interesting, most probably it has nothing to do with a realistic cosmology.

There is quite a long list of papers where the attempts has been made to solve
the problem of the cosmological constant along similar lines. A list of
references, probably non-complete, can be found in paper \cite{ad3}. All these
attempts proved to be not successful. There is even a no-go theorem \cite{sw1}
which states that a scalar field cannot successfully solve the problem of
adjustment of vacuum energy. Because of that we will turn in the next section
to fields with higher spins.

\section{Vector Field and the Adjustment Mechanism}

At a first glance a condensate of vector or higher rank tensor field would
destroy the observed homogeneity and isotropy of the universe and that is why
the earlier attempts to realize adjustment of vacuum energy down to zero, were
based on a scalar field. This is not true however, because
space-point independent time components of
these fields as well as isotropic components of symmetric tensor field,
$S_{ij} \sim \delta_{ij}$, break neither homogeneity nor isotropy. Such
condensates would destroy of course Lorents invariance of the theory but
since such a field interacts with matter only gravitationally,
the breaking of Lorents invariance
would be at the same level as that induced by a choice of a preferable
cosmological frame where cosmic microwave radiation is isotropic, and is not
dangerous from the point of view of experiment. A theory of higher rank
tensor fields opens reacher possibility than that of just massless scalar field
and in particular presents a counterexample to the "no-go" theorem mentioned in
the previous section.

In ref. \cite{ad4} a gauge vector field with the usual kinetic term, like in
the Maxwell electrodynamics, $F^2_{\mu\nu}$, was considered.
Such field is stable in the De Sitter background. To induce an instability
the coupling to the curvature, $\xi R U (A_\mu^2)$, which breaks
gauge symmetry was introduced. The model contains too much
arbitrariness, connected with the choice of the potential $U(A^2)$,
and gives rise to a time dependent gravitational constant
though the dependence can be much milder than in the scalar case, e.g.
$G_N$ may logarithmically depend on time.

A more interesting model is based on the gauge non-invariant Lagrangian of the
form \cite{ad5,ad3}:
\be{
 {\cal L}_0 =  \eta_0 A_{\alpha;\beta}A^{\alpha;\beta}
\label{la}
}\ee
which contains only a simple kinetic term without any potential terms.
The classical equation of motion for the time component $A_t$ in this case has
an unstable solution and with the proper sign of the constant $\eta_0$
the energy-momentum tensor corresponding to this solution compensates the
vacuum one. Indeed the equations of motion for the field $A_\mu$ in
metric~(\ref{ds2}) have the form:
\be{
(\partial^2_t -{1\over a^2} \partial^2_j +3H \partial_t -3H^2) A_t
+{2H \over a^2} \partial_j A_j = 0,
\label{d2tat}
}\ee
\be{
(\partial^2_t -{1\over a^2} \partial^2_j +H \partial_t -\dot H - 3H^2) A_j
+2H \partial_j A_t = 0
\label{dt2aj}
}\ee

The energy-momentum tensor of this field is easily calculated from the
Lagrangian (\ref{la}) and is equal to:
\begin{eqnarray}
\eta_0^{-1} T_{\mu\nu} (A_\alpha) =
-{1\over 2} g_{\mu\nu} A_{\alpha;\beta}A^{\alpha;\beta} +
 A_{\mu ;\alpha}A_\nu^{;\alpha} + A_{\alpha;\mu}A^\alpha_{;\nu} -
\nonumber \\
{1\over 2}\left( A_{\mu;\alpha} A_\nu +A_{\nu;\alpha} A_\mu +
A_{\alpha;\mu} A_\nu + A_{\alpha;\nu} A_\mu -
A_{\mu;\nu} A_\alpha - A_{\nu;\mu} A_\alpha \right)^{;\alpha}
\label{taa}
\end{eqnarray}
The Hubble parameter which enters equation (\ref{d2tat}) is determined by the
expression:
\be{
3H^2 m_1^2  = \rho_{tot} = \rho_{vac} + T_{tt}
\label{hat}
}\ee
where $m_1^2 = m_{Pl}^2/8\pi$.

We will consider a special homogeneous solution: $A_j =0$ and $A_t =A(t)$. We
assume that initially the magnitude of $A_t$ is small and the expansion of the
universe is dominated by the vacuum energy,
$H_v = \sqrt{ 8\pi \rho_{vac} /3m_{Pl}^2}$. In this regime $A_t$ exponentially
rises, $A_t(t) \sim \exp (0.79 H_vt)$ and soon its contribution into the energy
density becomes non-negligible. If $\eta_0 = - 1$ is chosen, so that
the vacuum energy
density and the energy density of the field $A_t$ has opposite signs, the
contribution of $A_t$ would diminish $H$ and both the expansion rate
and the rate
of increase of $A_t$ would slow down. One can check that asymptotically
$A_t \sim t$ and $H=1/t$. Expanding the solution in powers of $1/t$ and
assuming that $\rho_{vac}> 0$ and $\eta_0 = -1<0$  we find:
\beq{
A_t = t\sqrt{\rho_{vac} /2} \left( 1 + {c_1 \over t} + {c_2 \over t^2}
\right)
\label{aasym}
}\eeq
\beq{
H = {1\over t} \left( 1 - {c_1 \over t} + {c_1^2 - 4c_2/3 \over t^2}\right)
\label{hasym}
}\eeq
where $c_2 = 3m_{Pl}^2/8\pi \rho_{vac}$ and $c_1$ is determined by initial
conditions. The energy and pressure density of this solution are respectively
\beq{
\rho(A_t) = {1\over 2}\dot A_t^2 + {3\over 2} H^2A^2_t \rightarrow
\rho_{vac} (-1 + c_2/t^2)
\label{rhoat}
}\eeq
and
\beq{
p(A_t) = \rho_{vac} (1 - c_2/3t^2)
\label{pat}
}\eeq

>From eq.(\ref{rhoat}) we obtain the following expression for the Hubble
parameter:
\beq{
H^2 = {\rho_{vac} + \eta_0 \dot A_t^2/2 + \rho_{matter}
\over 3(m^2_1 -\eta_0 A_t^2/2)}
\label{h2}
}\eeq
The energy density of normal matter, $\rho_{matter}$, is added
here for generality. Since
$\rho_{matter} \sim 1/a^4$ for relativistic matter, $\rho_r$,
and $\sim 1/a^3$ for non-relativistic matter, $\rho_{nr}$, the contribution
of the usual matter into total cosmological energy density quickly dies down,
$\rho_r\sim 1/t^4$ and $\rho_{nr}\sim 1/t^3$, and becomes negligible. Thus the
result $H\ = 1/t$ does not depend on the matter content and follows from the
asymptotic rise of the field, $A_t \sim t$. The total
cosmological energy density in this model is dominated by the remnant of
$(\rho_{vac}-\rho_A )\sim 1/t^2$. This cosmology is not realistic and
in particular because
the expansion rate, $a(t) \sim t$ is too fast. One can try to construct a model
with a slower expansion rate using the freedom of adding new derivative terms
into the Lagrangian:
\beq{
{\cal L}_1 =\eta_1 A_{\mu;\nu}A^{\nu;\mu}
\label{l1}
}\eeq
\beq{
{\cal L}_2 =\eta_2 (A^\mu\>_{;\mu})^2
\label{l2}
}\eeq
However the first one gives exactly the same equation of motion for $A_t$ as
the Lagrangian ${\cal L}_0$ and the contribution from ${\cal L}_2$ into
the equation of motion is just $ \eta_2 A^\alpha_{\>;\alpha;\mu}$. It does not
change the asymptotic behavior obtained above.
So for a more realistic cosmologies one
has to address higher rank fields. We will do that in the next section.

Let us consider now the contribution of the space components $A_j$ into the
energy density. It follows from eq. (\ref{dt2aj}) that in the cosmological
background with $H=1/t$ the space components
$A_j$ increase as $t^{\sqrt 2}$ i.e. even faster
than $A_t$, but the energy density of these components remain small in
comparison with $\rho (A_t) \approx const$ (\ref{rhoat}):
\beq{
\rho (A_j ) =  {1\over a^2} \left( -{1\over 2} \dot A_j^2 + H \dot A_j A_j
-H^2 A_j^2 \right) \sim t^{2\sqrt{2} -4} = t^{-1.17}
\label{rhoaj}
}\eeq
However since $\rho (A_t)$ is canceled with $\rho_{vac}$ down to terms of the
order $1/t^2$, the contribution of $\rho (A_j)$ becomes dominant. Moreover
the energy-momentum tensor of $A_j$ contains undesirable non-isotropic terms
proportional to $A_iA_j$ or to $\dot A_i A_j$.
These terms can be suppressed if one adds the
Lagrangian ${\cal L}_1$ (\ref{l1}) with the proper choice of parameter
$\eta_1$.
One can check that in this case the space components rise as
$A_j \sim t^{\sqrt{ 2(1+\eta_1/\eta_0)}}$. So for $-1<\eta_1/\eta_0 <-1/2$ the
contribution of $A_j$ into cosmological energy density would be small. Though
the model of this Section is not realistic the tricks used here may be useful
for more realistic models considered in the following Section.

One more comment about the cosmological solutions with $A_t$ may be of
interest.
Let us assume now that $\eta_0$ is positive, $\eta_0 = 1 $. Corresponding
cosmological model in this case possesses a rather peculiar singularity.
The equation of motion
(\ref{d2tat}) does not change and the field $A_t$ remains unstable in the
Robertson-Walker background but the behavior of the solution becomes quite
different. One can see from eq. (\ref{h2}) that the Hubble parameter $H$ has
a singularity during expansion stage at a finite value of the field amplitude
and at a finite time. The solution near the singularity has the form:
\beq{
H = {h_1 \over (t_0 - t)^{2/3} },
\label{hsin}
}\eeq
\beq{
 A_t(t) =  \sqrt 2 m_1 \left[1 + c_1 \left( t_0 - t\right)^{2/3} \right]
\label{atsin}
}\eeq
where $m_1 = m_{Pl} /\sqrt{8\pi}$ and $c_1$ and $h_1$ are constant.
The energy density of the field $A_t$ at the singular point
tends to infinity as $(t_0 - t)^{-2/3}$
while the scale factor tends to a constant value according to the expression
$a(t) \sim \exp [-3h_1 (t_0 -t )^ {1/3}]$.

\section{Second Rank Symmetric Tensor Field}

Essential features of cosmologies with higher rank symmetric tensor fields are
the same as discussed in the previous section
but some details may be different and in particular the expansion
rate. Equation of motion for the space-point independent
components of the second
rank symmetric tensor $S_{\alpha\beta} $ in the flat RW background (\ref{ds2})
has the form:
\beq{
 (\partial_t^2 + 3H\partial_t -6H^2) S_{tt} -2H^2 s_{jj} = 0
\label{dt2tt}
}\eeq
\beq{
 (\partial_t^2 + 3H\partial_t -6H^2) s_{tj} = 0
\label{dt2tj}
}\eeq
\beq{
 (\partial_t^2 + 3H\partial_t -2H^2) s_{ij} -2H^2 \delta_{ij} S_{tt} = 0
\label{dt2ij}
}\eeq
where  $s_{tj} = S_{tj}/a(t)$ and $s_{ij} = S_{ij}/a^2(t)$.

For $\eta_0 = -1$ there exists a particularly interesting homogeneous
solution of these
equations which at large $t$ behaves as $S_{tt} = Ct$, $s_{ij} =
\delta_{ij} Ct /3$, and $s_{tj} =0$. The condition of vanishing of $s_{tj}$
is not stable but its stability can be ensured in the same way as stability
of space components $A_j$ discussed in the previous Section.
There may be non-vanishing components $s_{ij}$, which are not proportional
to the isotropic tensor $\delta_{ij}$, but they rise with time slower
than $t$. The energy density corresponding to this solution
\beq{
 \rho = \eta_0\left[{1\over 2} (\dot S^2_{tt} + \dot s^2_{ij} )
+ H^2 (3S^2_{tt} + s_{ij}^2 + 2 S_{tt}s_{jj} )\right]\,,
\label{rhostt}
}\eeq
exactly compensates the vacuum energy density, as above in the case of vector
field, but the expansion rate at large $t$ is different:
\beq{
 H = {3\over 8t}
\label{3h8t}
}\eeq
In this model $a\sim t^{3/8}$ and the energy density of usual matter decreases
rather slowly, $\rho_r \sim t^{-3/2}$ and $\rho_{nr} \sim t^{-9/8}$.
Corresponding values of the parameter $\Omega = \rho_{matter} /\rho_c$ would
be much larger than 1. Though the energy density of the usual matter may be
the dominant one, the Hubble parameter, as above,
 does not depend on it. Using expression
(\ref{rhostt}) we find similarly to (\ref{h2}):
\beq{
 H^2 = { \rho_{vac} + \eta_0 (\dot S_{tt}^2 + \dot s_{ij}^2) /2 +\rho_{matter}
 \over 3 m_1^2 - \eta_0 (3S_{tt}^2 + s^2_{ij} + 2S_{tt}s_{jj} ) }
\label{h22}
}\eeq
One can easily check that the asymptotic solution of the equation of motion
$S_{tt}\sim t$ and $s_{ij} \sim t$ gives the result (\ref{3h8t}) independently
of the energy density of the usual matter, $\rho_{matter} $,
and its equation of state. This is in a drastic contrast to the standard
cosmology, when the expansion rate is determined by the usual matter, so for
an agreement with observations a particular fine-tuning is necessary even if
one manages to obtain a normal expansion rate. To achieve the latter we can use
the freedom in the choice of the Lagrangian of the tensor field similar to
expressions (\ref{l1}) and (\ref{l2}):
\beq{
\Delta {\cal L} = \eta_1 S_{\alpha\beta;\gamma} S^{\alpha\gamma;\beta}
+\eta_2 S^\alpha_{\beta;\alpha} S^{\gamma\beta}_{\> \>;\gamma}
+\eta_3 S^\alpha_{\alpha;\beta} S_\gamma^{\gamma;\beta}
\label{deltal}
}\eeq

Varying the coefficients $\eta_j$ one can get different expansion regimes,
in particular $a(t) \sim t^{1/2} $ or $a(t) \sim t^{2/3} $ which correspond
respectively to radiation domination (RD) and matter domination (MD)
expansion regimes in the standard cosmology.
However this cosmological scenario has a serious problem which is related to
the fact that the Hubble parameter does not depend upon the equation of state
of cosmic matter. In particular it is not clear how to change the regime from
RD to MD, how to satisfy the nucleosynthesis constraints, and many other
constraints on the way of constructing a realistic cosmology.

Except for more freedom for realization of different expansion regimes
there is one more advantage of symmetric tensor field in comparison to the
vector one. Namely, quantum corrections for a non-gauge field generically
induce a nonzero mass even if one has started with a massless theory. In the
case of considered above vector field no principle is seen which could prevent
from the quantum mass generation. For the symmetric tensor field there is
symmetry of the Lagrangian with respect to transformation
$S_{\mu\nu} \rightarrow S_{\mu\nu} + C g_{\mu\nu}$ with an arbitrary constant
$C$. This symmetry does not permit to generate mass terms by quantum
corrections.

\section{Conclusion}

The models described above have at least a partial success in solving the
mystery of vacuum energy. We started with dominated by
cosmological term De Sitter universe, which expanded exponentially, and came
to the Friedmann type universe, which expanded as a power of time. The
energy-momentum tensors of the classically unstable condensates of the vector
or tensor fields asymptotically tend to the vacuum one (with the opposite
sign), so that the vacuum energy is canceled out in accordance with our
expectations.

Unfortunately there are still quite many serious problems on the road to
a realistic cosmology. The most serious one is mentioned at the end of the
previous section, that the expansion regime is not related to the matter
content in the universe. It is determined by the new fields and in this new
cosmology there should be a new fine-tuning, which makes the density of the
usual matter so close  to the critical one.

If despite all these shortcomings the solution to the problem of vacuum energy
is achieved by the compensating field, it seems then that all such theories
have the generic feature that there is always a remnant of non-compensated
vacuum energy density which is close to the critical one. Correspondingly
at all stages of the universe evolution the role of cosmological constant
should be essential. It looks like a model with a time-varying cosmological
constant~\cite{ad2} which  always, during nucleosynthesis, at the onset
of structure formation, and now, is 100\% essential. In these frameworks the
problem is why the traditional cosmology describes observations so well.

There are also quite serious problems associated with the theory of
(massless) non-gauged vector or tensor fields especially in
its quantized version. Usually a theory
of high spin field is formulated in such a way that lower spin components
are suppressed by an additional condition. In the models which
are considered here the condition is opposite, namely we have to
exclude the highest spin component in the theory. It is an open question
if such a theory may be formulated in self-consistent way.
Still, keeping in mind the gravity of the cosmological constant problem, maybe
the price paid here for its possible solution is not too high.

\section*{Acknowledgments}
This work was supported in part by Danmarks Grundforskningsfond through its
funding of the Theoretical Astrophysical Center (TAC).

\newpage

\end{document}